# The Intrafirm Complexity of Systemically Important Financial Institutions*

*R.L. Lumsdaine[(a)], D.N. Rockmore[(b,f)], N. Foti[(c)], G. Leibon[(b,d)], J.D. Farmer[(e,f)]*

[(a)]Kogod School of Business, American University, National Bureau of Economic Research and Center for Financial Stability
[(b)]Department of Mathematics, Dartmouth College
[(c)]Department of Statistics, University of Washington
[(d)]Coherent Path, Inc.
[(e)]Institute for New Economic Thinking at the Oxford Martin School and Mathematical Institute, University of Oxford
[(f)]The Santa Fe Institute

May 8, 2015

In November, 2011, the Financial Stability Board, in collaboration with the International Monetary Fund, published a list of 29 "systemically important financial institutions" (SIFIs). This designation reflects a concern that the failure of any one of them could have dramatic negative consequences for the global economy and is based on "their size, complexity, and systemic interconnectedness". While the characteristics of "size" and "systemic interconnectedness" have been the subject of a good deal of quantitative analysis, less attention has been paid to measures of a firm's "complexity." In this paper we take on the challenges of measuring the complexity of a financial institution by exploring the use of the structure of an individual firm's *control hierarchy* as a proxy for institutional complexity. The control hierarchy is a network representation of the institution and its subsidiaries. We show that this mathematical representation (and various associated metrics) provides a consistent way to compare the complexity of firms with often very disparate business models and as such may provide the foundation for determining a SIFI designation. By quantifying the level of complexity of a firm, our approach also may prove useful should firms need to reduce their level of complexity either in response to business or regulatory needs. Using a data set containing the control hierarchies of many of the designated SIFIs, we find that between 2011 and 2013, these firms have decreased their level of complexity, perhaps in response to regulatory requirements.

Keywords: SIFI, control hierarchy, macroprudential regulation
JEL classification codes: G28, G21, G01, C02

* Acknowledgements: We would like to thank seminar participants at Dartmouth College and conference participants at the American University Info-Metrics Institute's Information and Econometrics of Networks workshop for helpful comments, discussions, and suggestions related to an earlier draft of this paper. Any errors are our own. Farmer, Foti, and Rockmore were supported in part by the Alfred P. Sloan Foundation; Foti was additionally supported by a grant from the Dartmouth College Neukom Institute for Computational Science. Lumsdaine is grateful for the generous hospitality during various visits to both Erasmus University Rotterdam and the University of Portsmouth (UK) in contributing to the completion of this project. Address correspondence to: robin.lumsdaine@american.edu

1. **Introduction**

The Financial Stability Board (FSB) describes a systemically important financial institution, or SIFI, as a financial institution "whose disorderly failure, because of their size, complexity and systemic interconnectedness, would cause significant disruption to the wider financial system and economic activity."[1] Developed in the aftermath of the recent global financial crisis, this characterization represents an expanded regulatory definition relative to earlier ones based primarily on size (e.g., the list of "mandatory banks" subject to the Basel II capital regulations, see 72 FR 69298, December 7, 2007).

In particular, the collapse of Lehman Brothers in September 2008 highlighted the extensive *inter*connectedness of the financial system and the importance of considering not just the risk of a single firm but the risk to the entire financial system, i.e., the systemic risk. Interconnectedness can be formulated mathematically in terms of *networks* and indeed, a May 10, 2013 speech by Fed Chairman Ben Bernanke noted that "Network analysis, yet another promising tool under active development, has the potential to help us better monitor the interconnectedness of financial institutions and markets." In fact, there are a number of studies applying the techniques and metrics of network science to the analysis of economic and financial networks (e.g., Hutchinson et al., 1994, Cohen-Cole et al., 2010, Haldane and May 2011, Adamic et al., 2012, Battiston et al., 2012, Hautsch et al., 2012, 2013, Squartini et al., 2013, Caccioli et al., 2014). These papers have shown that the network of interdependencies is complex, and dependences of many different types overlap and interact (May, Levin, and Sugihara 2008). Because financial institutions hold various levels of interest in one another (Vitali et al., 2011), the collapse of any single entity can initiate a cascade of unforeseen events that in the worst of cases brings about the failures of many other financial participants, be they individuals, institutions, or even sovereign nations (Foti, et al., 2013).

While the interrelationships among financial network participants now are being widely studied, there has been comparatively little development of metrics concerning the *complexity of the individual firms* that comprise the system – the other key attribute highlighted in the SIFI definition. Failing any direct definition, one view of an individual firm's complexity comes from the lens of governance: "high complexity" would be interpreted as a corporate control structure rife with governance challenges for a firm's management, resulting in a lack of oversight that in turn poses significant operational, reputational, and balance sheet risk (Vitali et al., 2011). For example supervisory challenges have been evident in news surrounding both JP Morgan Chase's large trading loss and Barclays PLC's LIBOR fine, with senior management at both firms denying knowledge of the underlying operational lapses. A similar "rogue trader" event occurred at Société Générale in July 2007 and threatened to disrupt financial markets before it was determined to be an isolated incident. These examples illustrate the challenges associated with assuring sufficient oversight in organizations that often have very complex control and governance structures. Such complexity contributes to the possibility that subsidiaries act in relative obscurity within the organization, in spite of their significance to the overall viability of the parent institution. In this context, complexity therefore poses risk to the organization. When coupled with a high degree of interconnectivity, the combination can pose a risk to the global financial system as a whole.

---

[1] www.financialstabilityboard.org/publications/r_101111a.pdf. Accessed 5/5/2011.

In addition to the challenges that a firm's management faces, complex control hierarchies also present difficulties for regulators tasked with supervisory oversight. Prior to the recent financial crisis, banking supervision largely focused almost exclusively on institution-specific supervision. Since then, a recognition of the insufficiency of this approach in preventing the crisis has led policymakers to instead emphasize "macroprudential policies [that] differ from purely microprudential approaches in that they are intended to protect the financial system as a whole and, by extension, the broader economy" (Yellen 2010), an effort that is behind the SIFI designation. Yet despite this stated goal, many of the systemic risk mitigation efforts by regulators have been implemented via aggregation of firm-specific information (e.g., call reports, stress tests) and the imposition of uniform reporting requirements. This aggregation has occurred despite substantial variation in the organizational structure and mix of business activities found in financial institutions. Greater complexity (in terms of organizational structure and business activities) of an individual firm in turn makes it harder for a supervisory entity to disentangle and understand the firm's larger systemic interconnectedness and increases the likelihood that some parts of the firm's activities and interrelationships go unnoticed. In the case of large multinational organizations, a complexity measure related to oversight would naturally account for the burdens posed by coordinating over multiple national and regulatory environments. For the supervisor and regulator, complexity and opacity are often synonymous. Therefore, the identification of metrics that enable comparison across firms that may have very different control hierarchies is of critical importance in the post-crisis environment.

One view of the organizational structure of a firm comes from its *control hierarchy*([Vitali et al., 2011). This consists of the (parent) company and all of its subsidiaries, considered in its natural hierarchical and networked arrangement. This is effectively a standard representation of the *intra*connectedness of a firm. While network complexity is a well-studied subject (see e.g., Bonchev and Buck 2005), much of this literature seems not to be applicable to the very specific kinds of network topologies of the control hierarchies, which are *rooted directed trees* (in the parlance of computer science). Figure 1 shows a simple example of such an object.

Relevant, but different, is the literature and work in the measurement of business firm complexity or operational complexity. For example, Grant et al., (2000) define "operational complexity in terms of its lines of business and geographic operations regions, considered separately, jointly, and interactively" and use it to consider how such complexity is associated with equity analyst following. They classify firms according to operating segments, a unit of measure that recognizes the link between a firm's geographic and business line structures. They argue that such an approach is more parsimonious (and hence may improve inference) than considering each dimension separately, and is particularly appealing for firms that are highly segmented in terms of which business lines operate in specific countries. As a result, they use the product of the number of business lines and geographic regions as a measure of a firm's operational complexity.

This paper explores the possibility of using network-based metrics to encode the organizational complexity of many of the SIFI institutions. Ours is a novel approach that uses the innate network structure of the control hierarchy. In doing so, we therefore highlight the importance of considering *intra*-firm complexity in addition to the more-commonly-studied *inter*-firm complexity (i.e., the interconnectedness across firms) when determining SIFI designations. As we explain below, we see this network representation of the control hierarchy, as well as the metrics we construct, as intimately related to the kinds of oversight/regulatory concerns that we have outlined above.

To this end, it is worth noting that in 2011, the Financial Stability Oversight Council (FSOC), tasked under Dodd-Frank with the authority to confer the SIFI designation, began doing so, much to the chagrin of some of the affected firms. At least one of these firms (Prudential) stated its intention to contest the designation.[2] Additionally, MetLife very publicly deregistered as a bank holding company and sold portions of its business in an attempt to avoid the SIFI designation and the more stringent capital requirements that come along with it. It is interesting, therefore, to consider the way in which organizational complexity may respond to regulatory oversight. As the implications of being designated a SIFI evolve, criticisms over the lack of quantitative metrics to evaluate the appropriateness of the designation have emerged.[3]

In addition to aiding our understanding of the firms' (hierarchical) complexity, we use the network framework to identify new metrics for operational risk in these market participants. Such metrics may be useful for the creation of an "early warning system" for the kinds of institutional opacity that might in turn feed into an early identification of increased systemic risk. The metrics are also designed to inform supervisory judgment regarding the SIFI designation. The network encoding and associated metrics also open the door for the use of simulations as a means of assessing changes in complexity should a firm alter its business structure via a change in control hierarchy. Such simulations could provide a helpful tool for understanding the supervisory implications of altering a firm's control hierarchy in the process of winding down a firm (such as in the case of the dismantling of Lehman Brothers), or in arranging a rapid acquisition, (e.g., in the cases of the JP Morgan Chase acquisition of Bear Stearns, the Wells Fargo acquisition of Wachovia, or the Bank of America acquisition of Washington Mutual). The goal would be to reduce, rather than increase, systemic risk in the wake of a crisis and these metrics provide a means of comparing the organizational possibilities.

Among the questions we address using network-based metrics are:
- Are SIFI institutions indeed more complex?

---

[2] See statement on Prudential's website: http://news.prudential.com/article_display.cfm?article_id=6608. Since the initial draft of this paper, Prudential has withdrawn its plan to contest the designation.

[3] See, for example, the Remarks by [then-]FDIC Chairman Sheila C. Bair – "We Must Resolve to End Too Big to Fail,", May 5, 2011, where she notes, "That's why it is important that the FSOC [Financial Stability Oversight Council] move forward and develop some hard metrics to guide the SIFI designation process." Available at http://www.fdic.gov/news/news/speeches/chairman/spmay0511.html.

- Do size rankings match complexity rankings that utilize country and SIC (Standard Industrial Classification) code information?
- How has the complexity of SIFIs changed over time?

## 2. Data

We use an anonymized data set provided to us by Kingland Systems[4] of twenty-nine large financial institutions that include 19 of the original 29 SIFIs and 10 other firms (5 non-SIFI banks and 5 insurance companies). See Appendix A for a complete list of the firms. In the analysis, the data are numbered in random order within group (i.e., SIFIs, non-SIFI banks, and insurance companies) to protect confidentiality.

For each firm, we obtain underlying data that encodes the control hierarchy. As described above, this is the evolving intra-institutional system of relationships that stems from the "ultimate parent" (the SIFI of interest) through the ongoing process of creation, acquisition, and dissolution of subsidiaries by various entities in the institution. In this context "control" is defined as an ownership stake of at least 51% and a seat on the board, but in actuality the articulation of control can be more subtle than that (e.g., it may depend upon the nature of ownership in terms of the kinds of interest – voting or non-voting).[5] In addition to the control relationships the dataset also contains the country of origin and SIC[6] (Standard Industrial Classification) code of each entity. We have data for the twenty-nine institutions at two distinct dates, May 26, 2011 and February 25, 2013.

## 3. Methods

We draw on techniques from the science of *networks* to analyze the organizational structure of these large market participants (see Newman 2010 for a basic reference). Network analysis has already proven important for its ability to articulate complex interrelationships in a host of other disciplines, especially social systems (e.g., social networks) (Watts 2004), ecological networks (Dunne 2006), a variety of biologically-based networks (Dodds and Watts 2005; Allesina and Pascual 2009), management and organizational structure (e.g., Mirrlees 1976; Cremer 1980), and of particular relevance, financial and economic networks (e.g., Thurner et al., 2003; Boss et al., 2005; Leibon et al., 2008; May et al., 2008; Kyriakopoulos et al., 2009; Cohen-Cole et al., 2010; Jackson 2010; Tumminello et al., 2010; Adamic et al., 2012; Avraham et al., 2012).

The networks describing the organizational control structures of the firms in our study are characterized by a rooted directed tree structure. This type of network is composed of *nodes* and (directed) *edges* (see Figure 1). A *tree* is a network without loops. It is *directed* if the links come with a direction. Note (as indicated in Figure 1) a rooted tree has a preferenced node, the "root." In the trees of interest (representing control hierarchies) all edges point away from the root. An edge pointing from node A (the "parent") to node B (a "child") encodes the fact that entity A "controls" entity B (i.e., entity B is a subsidiary of entity A). The root of the tree is also called the *ultimate parent*. Nodes that have no directed edges to other nodes (i.e. they do not control any entities) are called *leaves*. The maximum number of nodes that a path from the root to any leaf would pass through is referred to as the *depth* of the tree. The number of children for a given node is called the *degree* of the node.[7]

---

[4] Kingland Systems is one of the leading companies that collects entity data, and specifically legal entity identification (see http://www.kingland.com/ for more information). We are grateful to Kingland, and especially to Tony Brownlee, George Suskalo, and Kyle Wiebers for their generosity in providing the data and their patience in answering our various questions.
[5] This definition was provided to us by Kingland Systems.
[6] https://www.osha.gov/pls/imis/sic_manual.html

Regulatory constraints create conditions of control that are reflected in the organization of the rooted trees derived from the control hierarchy. For example, the Banking Act of 1933 (more commonly known as the Glass-Steagall Act, after its Senator and Congressman sponsors) required a separation between commercial and investment banks (as well as other restrictions). In a network theoretic framework, this would mean that under Glass-Steagall, commercial and investment banks would need to be in different subtrees in the control hierarchy (e.g., in Figure 1, investment banks would need to be on one side of the root and commercial banks on the other). Similarly, legal and tax incentives might drive the patterns of country-incorporation, resulting in a tree structure where nodes associated with a specific business classification are also associated with a specific country. The 1999 passage of the Gramm-Leach-Bliley Act repealed many of the Glass-Steagall restrictions, fostering substantial growth-by-acquisition in the banking sector as banks diversified into new industries and countries. In tree terminology, this means that SIFI trees are no longer characterized by country- or SIC-specific subtrees but instead have become more jumbled. As the recent financial crisis unfolded, many viewed the repeal of Glass-Steagall as partially responsible, and calls to reenact it intensified, resulting in the 2010 Dodd-Frank Act. As a result, we might expect to see SIFI trees moving back toward their pre-1999 subtree separation. Therefore, the patterns of SIC and country codes (as node labels) in the control hierarchy help to reify regulatory changes and should be of interest to regulatory supervisors.

The tree structure of a SIFI is driven by basic conflict-of-interest considerations. That is, as we navigate through a chain of subsidiaries, we will never loop back to any of the intermediate (or parent) entities. As explained in Section 2, the control hierarchy imposes a tree structure on the data. For illustrative purposes, it is useful to select a firm with a small number of nodes to explore the information in the tree layouts further. Figure 2 considers SIFI S11; in 2011, this firm had 43 nodes, corresponding to 14 countries, four 1-digit SIC codes, and seven 2-digit SIC codes, with a tree depth of 4. In figures such as this, the largest circle represents the ultimate parent, with the size of other circles decreasing with growing distance from this parent; the smaller the circle, the farther down the tree it is. Figure 2a shows the layout of S11 by depth. Note that SIFI S11 has most of its subsidiaries at depth 3, with 28 entities distributed among four subsidiaries at depth 2. In addition, all but one of the children of the root (ultimate parent) is a leaf (does not control any additional subsidiaries). Thus most of the control hierarchy in this tree emanates from one child of the ultimate parent; in the absence of that one node, the tree would have depth 1. It should be evident, therefore, that severing the link between the root and that particular node would dramatically change the tree configuration, whereas severing the link with any other nodes at depth one would hardly change the configuration at all. Put another way, that node is central to the complexity of the firm's control hierarchy. Hence it may be a node of particular focus for a supervisor charged with evaluating the complexity of the SIFI.

---

[7] More generally with directed trees this is often referred to as the "out-degree" of the node, to distinguish between the "in-degree", that is, the number of immediate parents of a given node. However, in a rooted tree, all nodes except for the root have in-degree one, so throughout this paper we will use "degree" to refer to out-degree.

Figure 2b considers S11's control hierarchy when labeled by country. Despite the parent company being incorporated in Japan, most (five of eight) of its immediate children are incorporated in Great Britain, with only two incorporated in Japan and one in Greece. Yet among the 35 other children in the tree, all except the two US subsidiaries at depth 4 have an immediate parent that is also incorporated in Japan, suggesting from the perspective of a Japanese supervisor, it will be relatively easy to obtain information at all levels of control of the firm since there are only two entities beyond the immediate reach of the supervisor, assuming that a supervisor has access to all information within its own country.[8]

Figure 2c illustrates S11's control hierarchy labeled by 1-digit SIC code. At the 1-digit level, this firm is fairly homogeneous, with 31 of the children of the ultimate parent operating in the same industry. In addition to financial services, this firm has control over one entity in SIC area 3 (roughly construction and equipment), and 11 in services (areas 7 and 8); for the most part the services are concentrated in one subtree (to the left of the diagram). With most of the tree falling into the same SIC classification, it is evident that a financial services supervisor would be able to assess most of the firm's activities without having to rely on coordination with other supervisors. In addition, since the subsidiaries that fall under SIC classification 7 are concentrated in one subtree, the diagram highlights a single link that might warrant additional scrutiny or that might be severed should either the firm or supervisor wish to reduce the range of the firm's business activities. Figure 2d analogously illustrates S11's control hierarchy labeled by 2-digit SIC code.

For comparison, consider Figure 3 in which we see the same kinds of snapshots (at the same date) for a much larger firm (S16): this firm has 1778 nodes, corresponding to 32 countries and 100 SIC codes, with a tree depth of 5. Once again, we provide three representations color-coded according to distance from the root, country of origin, and 1-digit SIC code, respectively. All layouts were done using the freely available Gephi software package.[9] The layouts are normalized to be consistent across figures – that is, we have created layouts in which nodes are in the same positions from figure to figure. In all figures, the size of the circle represents tree distance from the root node. Figure 3a provides further detail and shows that the majority of nodes are at depth 2. Specifically, there are 299 nodes at depth 1, 1186 nodes at depth 2, 188 nodes at depth 3, 24 nodes at depth 4 and 80 nodes at depth 5. Figure 3b illustrates that in addition to the parent company being located in the US (the largest circle in the center), the majority of S16's subsidiaries are also located in the US. In addition, some of those US subsidiaries themselves have quite elaborate control hierarchies, judging from the large clusters at the top of the figure, as well as the ones on the far left, far right and bottom of the figure, that are almost exclusively comprised of US subsidiaries. There are also a large number of subsidiaries located in the UK (green), Germany (yellow), and Spain (light blue), with the UK and Spanish subsidiaries having a number of children that are located in the same country as their immediate parent. Figure 3c shows that while many of S16's subsidiaries are in the same 1-digit SIC classification as the parent company (classification 6 – Finance, Insurance, and Real Estate), the company also has a fairly diversified range of subsidiaries with the second largest 1-digit classification in the wholesale and retail trade sectors (SIC code 5).

---

[8] In some countries, for example the US, there are multiple regulators, so that even within-country coordination may be challenging. For the purposes of our discussion, we ignore this additional layer of complexity at the country level; it is reflected in the discussions of SIC complexity.
[9] https://gephi.org/

Taken together, Figures 2 and 3 illustrate that this view of complexity is multidimensional, that is, a firm that has a complex SIC classification structure may not have a complex country structure and vice versa. Identification of these complexity dimensions provides a way to compare sometimes very disparate firms and can inform allocation of supervisory resources aimed at managing systemic risk. In addition, it is necessary to develop metrics with which to compare firms' complexity.

### 3.1. Statistical Description of Heterogeneity

This section of the paper delineates some common metrics that quantify several important characteristics of the institutions' control hierarchy trees. These highlight the degree of heterogeneity in the firms' structures and provide a basis for comparison despite these differences.

A fundamental quantitative descriptor for any network (tree or not) is its *degree distribution*. This describes the probability distribution associated with the tree's set of degrees (i.e., the function *d(i)* that records the fraction of nodes with *i* children). Note that a tree with "regular branching" (i.e., where each non-leaf node is the parent of a constant number of children) produces a degree-distribution that is concentrated at two values, zero for any leaf node and some fixed non-zero constant value for all nodes that have children. For example, the rooted tree in Figure 1 has regular branching in which all nodes with children have out-degree two. Just as the degree distribution of a tree describing a firm's reporting lines might be used to characterize the spans of control (Urwick 1956) of its management, the degree distribution of a firm's control hierarchy analogously might be used to describe a supervisor's span of control in assessing the firm's systemic risk. Yet it is not just the *number* of children emanating from a node that is important for supervision. Rather it is the ease by which a supervisor can access information in order to make his or her assessment. For instance, an entity with many children that all fall under the same supervisor might be easier to assess than an entity with fewer but where the children in the latter case all fall under different country or SIC classification and hence require coordination across a number of supervisors.

For this reason, we therefore introduce a new metric that we believe is more closely linked to the supervisory interpretation of complexity that we have adopted, specifically a metric of complexity that we see as related to challenges in supervision derived from the need to coordinate oversight efforts across a variety of jurisdictions and agencies. More precisely, we assume that from a supervisory perspective, the simplest structure a firm can have is one where the firm and all its subsidiaries are situated in the same country and operate in the same industry. In such a setting, coordination burden is minimized. This forms the baseline for our analysis, and in assessing the complexity of any given firm, we derive a measure reflecting the distance of the control hierarchy tree from this "ideal" supervisory or *perfect tree*[10] structure. To this end we introduce the notion of *perfect tree similarity*.

In our framework, a perfect tree is comprised of *perfect groupings*, that is, country-specific (or SIC-specific) groupings that do not involve other countries (SIC codes). Our working hypothesis is that the greater the number of perfect groupings in a tree, the less likely that difficulty in one country (industry) will spill over into other countries (industries) in which the firm operates. Thus by comparing the firm's actual organizational structure to a perfect tree structure, we can draw inferences about the extent of contagion that a firm would experience were one of its subsidiaries to experience deterioration.

For each firm's baseline, we consider a "perfect," supervisory tree to be one with topology equal to the firm in question, where beginning with the nodes at level 2 each child has the same label (either country or SIC classification) as its immediate parent, -- i.e., we take as given the firm's heterogeneity at depth 1, reflecting its decisions regarding the distribution of business or geographic lines to each of its immediate subsidiaries (children). In this context the working hypothesis is that a "perfect" supervisory tree represents the simplest supervisory structure, one in which each supervisor's entities are purely domestic (country-perfect tree) or within a single industry (SIC-perfect tree). The underlying hypothesis is that less breadth of expertise is required to supervise/evaluate an entity that possesses a perfect supervisory tree (for example, across different national jurisdictions or industries) than in a firm that has a tree structure that is farther from perfect. Put another way, we accept each firm's prior business decisions regarding the distribution of countries (industries) in which it operates and are merely assessing the efficacy of its control hierarchy (organizational structure) from a systemic risk (supervisory) perspective within this distribution.

For any firm, we compare the proximity of their actual control hierarchy to their perfect supervisory baseline via a *perfect tree statistic,* that is, the fraction of nodes with the same label (i.e., country or SIC code) as their immediate parent. The statistic is therefore bounded between zero and one. Note that in a perfect tree (i.e., a tree in which the perfect tree statistic is equal to one), removal of a node and all subtrees below it will not change the value of the statistic.[11]  In contrast, a value of zero means that the subsidiaries below depth 1 are always different in character from their immediate parent (with respect to a given characteristic – country or SIC) and therefore would require maximal coordination among supervisors across all countries (industries) in which the firm operates. Thus, to the extent that a firm's tree structure is closer to a perfect supervisory tree, we reason that it is easier for both the firm and the supervisor to track/monitor the activities of its subsidiaries.

---

[10] That is, we begin by defining a "perfect tree" as one in which all nodes belong to the same (country or SIC) classification. In other contexts, a perfect tree refers to a tree with the same number of directed edges emanating from each node. It is important to recognize that our definition differs from that one.

[11] The supervisory analogue to this network structure might arise when a troubled firm is forced to sell or close one of its subsidiaries. The closer the organizational structure is to a perfect tree, the less likely there will be disruption to the rest of the firm when the subsidiary is pared.

The benefit of the perfect supervisory tree comparison is that it takes each firm's current organizational structure as given and does not require uniformity of business model across firms. Firms can be evaluated according to their own internal structure and the level of complexity they exhibit as a result of that structure can inform the allocation of supervisory resources. In this sense, we see this as a statistic that attempts to address the common "one-size-fits-all" criticism firms often make regarding the application of banking regulations to their business. The closer a firm is to its own perfect supervisory tree, the easier it will be for supervisors in different jurisdictions or industries to evaluate the portion of the firm they are responsible for without worrying about externalities that other parts of the firm might impose or contagion that a lapse in their oversight might cause to the rest of the firm.

In practice we recognize that the perfect tree characterization is too rigid and the costs associated with a fully-segmented structure may far outweigh the benefits of globalization and cross-border banking, but we believe it provides a useful regulatory ideal against which real world instances can be compared.

**3.2. Assessing tree design**

We evaluate the likelihood of a firm's organizational structure under the premise that more "common" designs (given a distribution of countries of incorporation/SIC classifications) might be easier for a supervisor to evaluate. Specifically, we take a firm's tree topology (degree distribution) as given and bootstrap all nodes and edges emanating from the ultimate parent, using 1000 replications, drawing from the firm-specific empirical distribution (i.e., with probability weights equal to those in the empirical distribution) of country and SIC codes. For each replication, *the perfect-tree-similarity statistic* is computed. The replications therefore allow us to generate a distribution of possible structures for each firm, conditional on its overall tree design (organizational structure), and to compute a range of summary statistics from these distributions. Such an approach allows a supervisor to compare supervisory complexity across firms that have different business models and country/industry profiles, while still holding the tree topology and distribution of countries/industries fixed for each firm.

**4. Results**

We present a large number of results describing the tree topology[12] of the banks. These are given in the following subsections:
1) Basic descriptive statistics
2) Degree statistics
3) Power law fits

---
[12] The "topology" of the tree or any network is the connectivity structure, that is, the layout of nodes and edges corresponding to the linkages in the network.

4) Parent-Child Transitions
5) Similarity statistics

We then use these metrics to draw inferences about the firms to determine whether there are differences between the different types of institutions. Finally we consider whether our metrics provide additional information beyond the size delineation that has traditionally been used to classify institutions that warrant greater regulation due to their systemic importance.

**4.1 Overall basic descriptive statistics.**

We begin with some basic descriptive statistics in Table 1, giving for each firm the total number of nodes (entities) in its tree, the number of distinct countries and SIC codes represented by the nodes, and the number of levels (i.e., depth) of each tree. The table illustrates the variation in tree structure across the firms and across time. In particular, in 2011 the number of nodes in a tree ranges from 43 to 16,443; the number of distinct countries and SIC codes represented by the nodes ranges from 10 to 89 and 13 to 281, respectively. In addition, the tree depth varies from 2 to 7. By 2013, there is less variation across firms. The number of nodes in a tree ranges from 330 to 12,752 while the number of distinct countries and SIC codes ranges from 23 to 86 and 27 to 164, respectively. The decline in number of countries and SIC codes is offset by the increase in tree depth for all but two firms, likely a result of post-crisis acquisitions. For some firms the increase in tree depth is substantial: for example, firm S17 increased from a tree depth of 4 in 2011 to 20 in 2013.

Table 1 is supplemented by Figures 4 and 5 which give a coarse scale indication of the overall structure of the dataset. The former shows the distribution of countries of incorporation over the totality of entities and the latter the distribution of SIC classifications at a granularity of one and two digits in the SIC.

**4.2. Overall degree statistics.**

One way to characterize complexity of the SIFI trees is via the hierarchy distribution, that is, the proportion of nodes at each level of the tree hierarchy. To aid our understanding of how tree hierarchies might be used in the context of large financial institutions, it is useful to first consider how different organizational structures correspond to different hierarchy distributions. For example, an institution with a very flat (i.e., "entrepreneurial") management structure would have a large proportion of nodes at level one and relatively few branches extending from those nodes. In contrast, an institution that concentrates its decision-making among only a few senior managers who are then held accountable for large portions of the firm would have a larger proportion of nodes at lower levels of the tree. Such a diffuse tree might also be found among organizations that have experienced significant growth by acquisition, such as many financial institutions in the decade preceding the recent financial crisis, where the tree of an acquired complex organization may have been grafted to the tree of the acquiring parent somewhere below the highest level, creating a very hierarchical structure of great depth (a "bureaucratic" structure).

Firms also might be arranged along geographical ("divisional") or industry ("functional") lines (Armour and Teece 1978).

The hierarchy distribution for our sample of firms is summarized in Figure 6, for 2011 and 2013 separately. Note that there is substantial variation across the firms. For example, while in 2011 more than one-third of the firms have more than half of their nodes at the first level of the tree hierarchy, others have relatively few nodes branching from the ultimate parent and instead have a large concentration of nodes farther down the tree (e.g., S13 and B1). None of the firms with the deepest trees (i.e., more than six levels) have node concentration at the first level of the tree hierarchy, indicating a flatter or more diffuse organizational structure. Note firm I4 has a tree structure that spreads out at each level in 2011. Across all firms in our sample, roughly one-third of the nodes are in each of the first two levels, another 22% in the third level, and only 10% at deeper levels in the tree hierarchy.

In contrast, by 2013, the tree hierarchies of the firms in our sample deepened substantially. For example, only half as many (five) firms now have more than half their nodes at the first level, while 14 have less than 10% of their nodes at the first level. In addition, 11 firms now have more than seven levels while just two years earlier, none did. Across all firms in the sample, by 2013 roughly 25% of the nodes were at deeper than the third level. Thus from the perspective of consolidated supervision, the challenges associated with assessing these firms increased dramatically, with many entities in the organization being much farther removed from the parent.

**4.3. Power law**

Of particular relevance for control hierarchies are the similarity of their degree distributions (described in Section 3.1) to so-called *power law distributions*. These are particular instances of heavy-tailed distributions and more specifically, are distributions that have the form $x^{-r}$ for some $r > 1$.

A wide variety of natural phenomena have been shown to exhibit behavior that can be approximated by a power law (Mitzenmacher 2004, Newman 2010). Networks with a power law degree distribution are also called *scale-free*, which is simply a shorthand for the property that the form of the degree distribution is such that it does not change under a simple scaling of the unit of measurement. Power law structures are consistent with a growth model of *preferential attachment* (Simon 1955). In this "rich get richer" scenario those entities that already have many subsidiaries are more likely to acquire new subsidiaries (in a manner proportional to their current fraction of subsidiaries). This is in contrast to a normal distribution, which would be consistent with a random growth model, or for that matter, a degree distribution concentrated or largely concentrated at a single value, that encodes a fixed branching. The latter scenario that might reflect some sort of internal rule about the number of subsidiaries that the firm believes should be controlled by a given parent. Table 2 contains power law results for all the firms in our sample and both time periods.

The heavy-tailed nature of the power law distribution is of particular interest vis-à-vis supervision. If there are a relatively few entities with an inordinate number of direct subsidiaries, then they would merit special attention and their relative scarcity would direct an efficient use of supervisory resources. Additionally, power law distributions have an interesting relation with the Gini coefficient (a commonly used measure of inequality): the higher the exponent (or rather the higher the absolute value of the exponent, since it is negative), the smaller the corresponding Gini coefficient of the distribution, so that the more "evenly distributed" (in the sense of Gini coefficient) are the nodes in terms of degree.[13]

**4.4 Markov Statistics (parent-child similarities)**

As noted above in the discussion of both degree and hierarchy distributions, from a systemic risk perspective it is not just the *number* (or proportion) of child nodes that emanate from a parent node that matters but also the *similarity* between the parent and child.[14] In this section we therefore ask the following: Given that a node is in a particular country A, what is the probability $P(A|A)$ that the node below it is also in country A? This is one simple measure of a type of "homophily" in the network, or the predilection for a node (in this case a parent) to be connected to another node of the same kind. See Easley and Kleinberg (2009), Chapter 4 for some discussion of this concept in the general network setting.

From a systemic risk perspective, such a metric is useful in identifying potential contagion effects should a firm start to exhibit signs of stress. The higher the in-country probability (and hence the lower the out-country probability), the more likely a supervisor will be able to contain disruption and avoid spillover effects. In Table 3 we compute this probability for each country, using all firms in our sample, for both 2011 and 2013. This varies dramatically for different countries. In 2011, Canada has the highest probability, with $P(A|A) = 0.97$; in contrast Switzerland has the lowest, with $P(A|A) = 0.11$. Part of the reason for this variation is differences in country frequency; for example under random assignment a country that has more nodes in the network has a greater likelihood of being paired with its own country than does a country that has fewer nodes in the network. By comparing the in-country probabilities in 2011 to those in 2013, however, we see that for most countries, in-degree probability increases, with many countries above 0.9. This suggests that firms have shifted their organizational structure to consolidate subsidiaries from a given country under parents from the same country.

---

[13] For a nice visualization see http://networkscience.wordpress.com/2012/04/19/power-law-paradox-power-law-exponent-does-not-mean-what-you-think-it-means/

[14] Although for expositional purposes much of our discussion has focused on the ease of a supervisor to "look below" in examining entities that are lower down the tree, our focus on the similarity between parent and child nodes stems from the view that risk management will be easier when, for example, a child has the same legal, accounting, tax, or supervisory framework as its parent.

## 4.5 Complexity and Changing Structure – Perfect Tree Similarity Statistics

In this section we document the fluidity of SIFI control hierarchy by comparing the perfect tree statistics in 2011 and 2013. As is the case for many complex systems, the structure of the control hierarchy is a response to a variety of endogenous and exogenous forces. A significant component of the latter comes from regulatory frameworks, which can include pressures that come from tax and corporate legal structures as well as supervisory restrictions on activities.

Table 4 summarizes the country-level perfect tree simulations for each firm and both dates, grouped by firm type (SIFIs, non-SIFI banks, and insurance companies). For each firm and each date, four statistics are reported: (1) the firm's "actual" perfect tree statistic (described in Section 3.2), (2) the mean of the same statistic computed for all 1000 replications of simulated data, (3) the standard deviation of the 1000 replications, (4) the empirical quantile corresponding to where the actual statistic would lie in the empirical distribution generated from the 1000 replications.

It is important to note that while perfect tree statistics of different values can be compared *within* a single firm (i.e., a lower value means more subsidiaries are different from their immediate parent than a higher value), care needs to be taken when comparing these statistics *across* firms because the proximity to a perfect tree depends on each firm's underlying topology. For this reason, it is useful to consider the moments and/or quantiles as a result of the perfect tree simulations. For example, a comparison of S9 and S11 in 2011 reveals that although S9 has a perfect tree statistic that is close to one, the arrangement of its subsidiaries is actually far from perfect (falling in the lowest third of the simulated distribution). In contrast, S11 has a much lower perfect tree statistic (0.279) but is in the 99.85% quantile compared to other control hierarchies that could result from its given country distribution. That means that given the distribution of countries at depth 1, its control hierarchy is "nearly perfect" in terms of ease of supervision, as most downstream child nodes belong to the same country as the corresponding node at depth 1.

In 2011, 9 of the 29 firms had country structures that were significantly different from their corresponding perfect tree, given their topology (that is, a test of the null hypothesis that the firm's perfect tree statistic is equal to one is rejected). In addition, 24 of 29 firms had country distributions that differed from a structure where all subsidiaries differ from their immediate parent (that is, a test of the null hypothesis that the firm's perfect tree statistic is equal to zero is also rejected), suggesting most firms have an organizational control structure that follows geographical lines.

Table 5 summarizes the SIC-level perfect tree simulations for each firm and both dates, grouped by firm type (SIFIs, non-SIFI banks, and insurance companies). Similar to the country tree statistics, while in 11 of the 29 firms we fail to reject the null hypothesis that they have SIC structures where all subsidiaries differ from their immediate parent in 2011 (perfect tree statistic is equal to zero), by 2013 the hypothesis is rejected for *all* of the firms' structures (at the 95% level of confidence). There was also a similar shift nearer to a perfect tree, as all except one firm's statistic increased and the number of firms with SIC tree-similarity significantly less than one fell from 20 to 10. Nonetheless, in many cases the change in SIC structure did not actually render the firms less complex, as the associated quantile declined in 14 of 29 firms, including all five of the non-SIFI banks.

An interesting example from this table is S18. From Table 1, we know that this firm experienced a modest (13%) reduction in nodes between 2011 and 2013, despite adding five additional countries to its control hierarchy. Over the same period, it reduced its number of SIC categories by more than 50% and doubled its degree depth. As noted above in Section 4.2, the increased degree depth is an indication of a shift toward a more bureaucratic organizational structure. Commensurate with this change, the firm's country perfect tree statistic increased both in level and quantile (see Table 4), indicating that the new control hierarchy was closer to a segmented structure, where children mimic the country of their immediate parent. This shift to a more divisional arrangement was not without cost, however; from Table 5 despite more parent-child SIC alignment (the perfect tree statistic increased from 0.666 to 0.844), the resultant firm was less functionally-arranged as it dropped from the $75^{th}$ percentile to the $37^{th}$ percentile in its proximity to a perfect tree.

## 5. Discussion

### 5.1. Are SIFIs really more complex from a supervisory perspective?

In Table 1, we see that in 2011 SIFIs had tree structures with more than three times as many nodes, higher degree depth, greater geographical reach and more than double the amount of SIC variation in their subsidiaries than both non-SIFI banks and insurance companies. In addition, SIFIs on average had $1.82tr in total consolidated assets as compared to $0.72tr for non-SIFI banks and $0.61tr for the insurance companies in our dataset. So do our complexity measures tell a regulator anything more about the appropriateness of a systemic risk designation than could be gleaned from simply looking at the size of an institution?

To reiterate, we consider the term "complexity" from a supervisory perspective, namely: (1) how difficult is it to supervise the firm, (2) how likely is a supervisor to identify a problem with the firm if it exists, and (3) how easily can a problem be mitigated/remedied once identified? We assume in this context that the closer a firm's control hierarchy to a perfect tree, the easier will be its supervision. This is because such a firm will require less coordination among supervisors in different geographical and industry jurisdictions. In other words, we assume the risk posed by institutions is related to the risk associated with an individual country/industry supervisor's ability to monitor the firms for which they are responsible, as well as the coordination across these various supervisors.

From this supervisory perspective, one might at first glance consider SIFIs to be on average more complex than either the non-SIFI banks or insurance companies, as a result of their much more elaborate control hierarchies based on the dimensions given in the data (# of nodes, countries, SIC groups, and degree depth. Yet for the most part they also had a greater proportion of child nodes that were from the same country or SIC classification as their immediate parent than did the insurance companies. Importantly, given their elaborate control hierarchy, they were in many cases closer to a perfect tree than most of the simulated trees, judging from their quantile position. Thus the SIFIs may not necessarily present greater supervisory challenges, despite their larger size and more elaborate structures, assuming sufficient oversight in each country and industry. Put another way, the ability of a firm to quickly reduce its exposure with respect to a specific country or industry appears to be similar between SIFI and non-SIFI banks. In contrast, insurance companies in 2011 appear to have more complex organizational structures, with more extensive cross-country and cross-industry reporting structures, that might prove harder to untangle in a crisis.

## 5.2. Is size a sufficient statistic?

In the aftermath of the recent financial crisis, calls to end "too big to fail," the so-called practice of bailing out the largest, most systemically important financial institutions, have intensified. Most often, the concept of too big to fail implies a firm whose size is larger than a specified threshold. Yet size is but one of the criteria mentioned in the SIFI definition.[15]

Consistent with the idea that size and systemic risk are synonymous, regulators typically delineate a size threshold to identify firms that pose considerable risk to the global financial system. Some recent examples include: (1) the Basel II capital regulations identified so-called "mandatory" adopters as "those with consolidated total assets (excluding assets held by an insurance underwriting subsidiary of a bank holding company) of $250 billion or more or with consolidated total on-balance-sheet foreign exposure of $10 billion of more" ( 72 FR 69290, December 7, 2007); (2) the Basel III final rule retained these threshholds and additionally articulated "enhanced disclosure requirements…for banking organizations with $50 billion or more in total consolidated assets", noting that small bank holding companies (those with total consolidated assets of less than $500 million) remained subject to a prior rule (12 C.F.R. 17, pts. 208, 217, and 225)[16]; and (3) the Dodd-Frank Act similarly articulated a $50 billion or more threshold to identify firms subject to a specific treatment of off-balance-sheet activities in capital computations.

Despite the ease of implementation, a size-based threshold is in many ways unsatisfactory, precisely because it does not take into account the level of complexity of a firm's business activities. To quantify this point more generally, we obtained data from Bloomberg® on total consolidated assets of all firms in our sample and computed the Pearson rank

---

[15] There are a variety of definitions of size that arise in the banking and finance literature. The most common in recent banking regulations (e.g., Basel II, Basel III, Dodd-Frank) is specified in terms of total consolidated assets. Other definitions might include market capitalization, number of distinct entities, number of employees. Generally speaking, however, in the "too big to fail" context, size is usually considered in financial (e.g., dollar) terms, rather than in terms of features of organizational structure.

[16] A number of other parts of the Basel III regulations were also based on a size threshold. For example, the length of allowable transition period for the phase-out of trust-preferred securities from tier 1 capital according to whether a firm had more or less than $15 billion in total consolidated assets (as of December 31, 2009),

correlation between that and both country- and SIC- perfect tree similarity statistics.[17] The rank correlation was -0.32 and -0.36, respectively. In contrast, the rank correlation between size and the number of nodes was 0.58, reflecting the fact that firms with more total consolidated assets generally have more subsidiaries. Taken together, these numbers highlight the fact that the link between asset size, number of subsidiaries, and the number of countries and industries in which a firm operates is not exact; that is, the perfect tree statistics are measuring something different than just the number of nodes. This point is also illustrated in Figure 7, comparing the ranks of the firms' size to the ranks of their country (top chart) and SIC (bottom chart) perfect tree similarity statistics. In particular, if size and the perfect-tree statistics measured the same thing, we would expect the points in this figure to lie along the 45-degree line. Instead there is a slight negative relationship, particularly among the SIFI firms; the larger firms are less complex from a supervisory standpoint, with a larger proportion of trees being self-contained (where the child node is in the same country as its immediate parent).

To see how such information might be used in practice, we offer two examples:

(1) Firms S4 and S5 have very similar asset-size, yet in 2013 S4 has three times as many subsidiaries (nodes) and activities in 50% more industries than S5. In contrast, it is only active in half as many countries. Hence the supervisory challenges associated with these firms may be very different, S4 requiring coordination among many more industry regulators and S5 requiring substantial coordination among different country supervisors. Yet both their country and SIC perfect tree-similarity statistics are similar and close to one. And while their perfect tree country statistics would place them at the 100% quantile (i.e., closer to a perfect tree than all 1000 simulated firms with identical tree structure), S4's SIC statistic, while higher in value than S5's, is below the 33% level (by comparison, S5's is just below the 54% level). Taken together, these statistics would alert regulators to the SIC dimension of S4's business. Note from Table 1 that the absolute number of SIC activities of S4 is not particularly unusual: there are 12 firms that have subsidiaries in a greater number of industries. Yet for its organizational structure, there is only one firm (S2) that has a statistic with a lower quantile score.

(2) Another comparison of interest is between S6 and S12. These two firms are very similar in their tree similarity statistics (both country and SIC), as well as the quantiles to which these statistics correspond. Both firms' SIC quantiles are near the median of the simulated distribution for their corresponding tree structures and hence might warrant additional supervision. Yet S6 is nearly 50% bigger in asset size and has more than four times the number of subsidiaries (nodes) than S12. A size-only threshold would potentially miss the complexity of S12.

---

[17] We report results using rank-based statistics in order to maintain confidentiality of the firms. The results are qualitatively similar using ln(assets).

**5.3. Has complexity changed?**

A comparison of the left and right blocks of columns in Table 4 and 5 (corresponding to results as of May 26, 2011, and February 25, 2013, respectively), reveals a number of interesting observations:

- In 2011, 6 of the 29 firms had country structures that were not significantly different from a random tree structure (a perfect tree similarity statistic of zero). By 2013, all of the firms' structures differed.
- Despite their nonrandom structure, in the 21 months between May 2011 and February 2013, the firms in our sample substantially reduced their level of geographical complexity. In the earlier sample, 9 firms (6 SIFIs, 2 non-SIFI banks, and 1 insurance company) had country structures that were statistically significantly different from a perfect tree (as perfect tree similarity statistic of one), while by 2013, only 1 remained.
- The reduction in geographical complexity was partially achieved through a change in structure and reduction in geographical intraconnectedness for each firm. For 25 of the 29 firms the country perfect-tree similarity statistic moved closer to one. In addition, over all 29 firms the range of statistics narrowed (from [0.261 to 0.963] in 2011 to [0.510 to 0.991] in 2013).
- By 2013, most firms were in the very upper tail of the country perfect tree distribution, with more than half the statistics above *all* of the simulated values corresponding to their firm's structure (quantile = 100%).
- In contrast, while most firms' SIC structure moved more closely to a perfect tree in 2013 (i.e., the number of firms with SIC perfect tree-similarity statistic significantly less than one fell by 50%, from 20 to 10), the associated quantiles declined in almost half the firms, including all five of the non-SIFI banks, indicating that the change in SIC structure did not commensurately reduce the firms' complexity along this dimension.
- As noted above, supervisors can also use measures such as these to inform their choices of which firms merit additional scrutiny. For example, for firm S2, despite an increase in its perfect tree similarity statistic between 2011 and 2013, the associated quantiles declined, indicating that the firm's geographical complexity may not have declined commensurately by the change in structure.

Taken together, these results indicate some reduction in country and SIC complexity for most of the firms in our sample. While SIFIs have made the most progress, there have also been large changes in the non-bank SIFIs and insurance companies. There is little evidence that the increase in size as a result of the post-financial crisis consolidation of the banking sector has led to greater jurisdictional complexity; for most firms, both supervisory oversight and possible wind-down or paring of assets would be easier now, given their organizational structure, than in 2011.

In contrast, an examination of the power law results indicates that most firms' power law exponents have increased, implying a lower Gini coefficient, that is, the degree distribution is more evenly distributed. Hence the firms in our sample appear to have reduced their country and SIC complexity while increasing their degree distribution complexity. What the implications of this shift are for systemic risk remains to be seen. On the one hand, it means that on average, each firm's risk is more evenly distributed across a variety of countries and industries, so that the control hierarchy is less highly concentrated in one country or industry, suggesting greater diversification of risk. On the other, it means there is a greater need for international and industry coordination among supervisors. In contrast, three of the non-SIFIs saw a reduction in their power law exponent, suggesting increased concentration. Such a move might represent a firm's decision to focus on its core business but it could also signal a greater sensitivity to business cycle fluctuations as a result of excessive concentration. These interpretations are left for future research.

**Conclusions**

The 2008 financial crises highlighted the risks that large, multinational, complex, interconnected banks pose. Since then, debate concerning which firms warrant a SIFI designation or are "too big to fail" has led to a large amount of research into the complexity inherent in financial transaction networks. Yet little emphasis has been placed on the challenges that a firms' internal complexity presents to supervisors tasked with evaluating a firm's riskiness.

In this paper, we propose using a firm's control hierarchy as a proxy for such supervisory challenges. By defining complexity as a function of the firm's tree topology we demonstrate that complexity and size are not synonymous and thus warrant distinct mention in the SIFI definition. We additionally propose using a perfect tree statistic to quantify the ease of supervision in a number of dimensions: (1) the need to rely on coordination or information from supervisors from other countries, (2) the implications of severing the link to a subsidiary, and (3) the ability to assess how to wind down a firm through paring of subsidiaries with minimal risk. By comparing data from 2011 and 2013, we find that on average the ease of supervision along these dimensions improved.

Contrary to conventional wisdom, our results suggest that some of the SIFI-designated institutions may not pose greater supervisory challenge since their control hierarchy more closely resembles a perfect tree than some of the other firms in our sample. We find little difference between SIFIs and non-SIFI banks. In contrast, the insurance companies in our sample are more complex according to these three criteria, despite being smaller in size, having fewer subsidiaries, and being less geographically or industry-diverse than the banks.

**References.**

Table 1. Descriptive Statistics

| | May 26, 2011 | | | | February 25, 2013 | | | |
|---|---|---|---|---|---|---|---|---|
| | #Nodes | #Countries | #SIC | Depth | #Nodes | #Countries | #SIC | Depth |
| **SIFIs** | | | | | | | | |
| S1 | 1007 | 34 | 72 | 3 | 1519 | 57 | 60 | 4 |
| S2 | 887 | 40 | 133 | 3 | 1585 | 47 | 59 | 10 |
| S3 | 2568 | 55 | 210 | 5 | 4001 | 70 | 125 | 7 |
| S4 | 1897 | 37 | 72 | 4 | 12752 | 33 | 86 | 6 |
| S5 | 1034 | 42 | 122 | 3 | 4272 | 56 | 57 | 5 |
| S6 | 3221 | 87 | 210 | 5 | 7289 | 73 | 96 | 7 |
| S7 | 5850 | 58 | 198 | 4 | 5477 | 72 | 164 | 11 |
| S8 | 6483 | 68 | 157 | 4 | 9564 | 76 | 147 | 9 |
| S9 | 5502 | 48 | 194 | 5 | 8455 | 47 | 127 | 9 |
| S10 | 1815 | 35 | 222 | 7 | 4012 | 48 | 107 | 8 |
| S11 | 43 | 14 | 16 | 4 | 1468 | 23 | 34 | 5 |
| S12 | 53 | 18 | 13 | 2 | 1520 | 29 | 45 | 5 |
| S13 | 935 | 32 | 46 | 5 | 2224 | 32 | 46 | 5 |
| S14 | 9815 | 76 | 281 | 5 | 3243 | 56 | 152 | 13 |
| S15 | 9084 | 89 | 240 | 6 | 10211 | 86 | 127 | 9 |
| S16 | 1778 | 32 | 100 | 5 | 2545 | 50 | 86 | 11 |
| S17 | 2334 | 49 | 250 | 4 | 1117 | 38 | 104 | 20 |
| S18 | 11487 | 47 | 279 | 6 | 10077 | 52 | 134 | 12 |
| S19 | 16443 | 58 | 172 | 6 | 11231 | 61 | 114 | 7 |
| *Mean* | *4328* | *48* | *157* | *4.5* | *5398* | *53* | *98* | *8.6* |
| **Non-SIFI Banks** | | | | | | | | |
| B1 | 2678 | 19 | 72 | 4 | 2378 | 30 | 65 | 5 |
| B2 | 1998 | 20 | 110 | 4 | 9079 | 31 | 63 | 4 |
| B3 | 127 | 37 | 26 | 5 | 681 | 50 | 49 | 7 |
| B4 | 475 | 32 | 147 | 3 | 7006 | 29 | 53 | 6 |
| B5 | 205 | 28 | 34 | 3 | 387 | 29 | 42 | 5 |
| *Mean* | *1097* | *27* | *78* | *3.8* | *3906* | *34* | *54* | *5.4* |
| **Insurance Companies** | | | | | | | | |
| I1 | 793 | 40 | 48 | 5 | 1373 | 39 | 67 | 6 |
| I2 | 118 | 25 | 27 | 5 | 330 | 30 | 27 | 6 |
| I3 | 1564 | 74 | 154 | 3 | 2738 | 81 | 131 | 7 |
| I4 | 1752 | 54 | 98 | 4 | 2544 | 48 | 86 | 7 |
| I5 | 379 | 10 | 47 | 4 | 1254 | 33 | 67 | 9 |
| *Mean* | *921* | *41* | *75* | *4.2* | *1648* | *46* | *76* | *7.0* |

Notes to Table 1: Basic descriptive information and statistics on the control hierarchies for the twenty-nine institutions in our sample at two points in time: May 26, 2011 and February 25, 2013. *#Nodes* is the total number of nodes in the hierarchy; *#Countries* is the number of different countries that occur in the hierarchy; *#SIC* is the number of different SIC codes that occur in the hierarchy; *Depth* is the number of levels in the hierarchy tree.

Table 2: Power law results

|  | 26-May-11 | | | 25-Feb-13 | | | chg in exp |
|---|---|---|---|---|---|---|---|
|  | exponent | se | xmin | exponent | se | xmin |  |
| **SIFIs** | | | | | | | |
| S1 | 1.68 | 0.022 | 3 | 1.73 | 0.019 | 2 | 0.05 |
| S2 | 1.81 | 0.027 | 5 | 1.94 | 0.024 | 3 | 0.13 |
| S3 | 1.52 | 0.010 | 1 | 1.78 | 0.012 | 1 | 0.26 |
| S4 | 1.51 | 0.012 | 2 | 1.53 | 0.005 | 1 | 0.02 |
| S5 | 1.58 | 0.018 | 1 | 1.74 | 0.011 | 1 | 0.16 |
| S6 | 1.70 | 0.012 | 2 | 1.75 | 0.009 | 2 | 0.05 |
| S7 | 1.71 | 0.009 | 2 | 1.85 | 0.011 | 1 | 0.14 |
| S8 | 1.53 | 0.007 | 2 | 1.69 | 0.007 | 3 | 0.16 |
| S9 | 1.46 | 0.006 | 1 | 1.68 | 0.007 | 1 | 0.22 |
| S10 | 1.74 | 0.017 | 2 | 1.61 | 0.010 | 1 | -0.13 |
| S11 | 2.38 | 0.218 | 4 | 1.48 | 0.012 | 1 | -0.90 |
| S12 | 2.49 | 0.218 | 7 | 1.68 | 0.017 | 2 | -0.81 |
| S13 | 2.07 | 0.035 | 16 | 1.49 | 0.011 | 1 | -0.58 |
| S14 | 1.71 | 0.007 | 30 | 2.06 | 0.019 | 7 | 0.35 |
| S15 | 1.52 | 0.005 | 4 | 1.6 | 0.006 | 1 | 0.08 |
| S16 | 1.48 | 0.011 | 1 | 2.02 | 0.020 | 2 | 0.54 |
| S17 | 1.64 | 0.013 | 2 | 1.84 | 0.025 | 1 | 0.20 |
| S18 | 1.59 | 0.006 | 3 | 1.82 | 0.008 | 4 | 0.23 |
| S19 | 1.50 | 0.004 | 3 | 1.72 | 0.007 | 1 | 0.22 |
| *Mean* | *1.72* | *0.03* | *4.8* | *1.74* | *0.01* | *1.9* | *0.02* |
| **Non-SIFI Banks** | | | | | | | |
| B1 | 1.54 | 0.011 | 5 | 1.62 | 0.013 | 1 | 0.08 |
| B2 | 1.36 | 0.008 | 1 | 1.49 | 0.005 | 1 | 0.13 |
| B3 | 1.78 | 0.070 | 1 | 1.66 | 0.025 | 1 | -0.12 |
| B4 | 1.56 | 0.026 | 1 | 1.44 | 0.005 | 3 | -0.12 |
| B5 | 2.25 | 0.088 | 6 | 1.61 | 0.031 | 1 | -0.64 |
| *Mean* | *1.70* | *0.04* | *2.8* | *1.56* | *0.02* | *1.4* | *-0.13* |
| **Insurance Companies** | | | | | | | |
| I1 | 1.67 | 0.024 | 8 | 1.55 | 0.015 | 1 | -0.12 |
| I2 | 2.06 | 0.098 | 3 | 2.2 | 0.066 | 4 | 0.14 |
| I3 | 1.61 | 0.015 | 1 | 1.64 | 0.012 | 1 | 0.03 |
| I4 | 1.48 | 0.012 | 1 | 1.67 | 0.013 | 1 | 0.19 |
| I5 | 1.42 | 0.022 | 1 | 1.57 | 0.016 | 1 | 0.15 |
| *Mean* | *1.65* | *0.03* | *2.8* | *1.73* | *0.02* | *1.6* | *0.08* |

Notes to Table 2: This table shows the power law exponent and associated standard error (se) for each of the firms in 2011 and 2013, as well as the change in the exponent (chg in exp). xmin is the minimum (lowest) depth for which the power law applies. See Clauset et al (2009), equation 2.6.

Table 3: Within-country "birth" probabilities P(A|A) aggregated over all firms in our sample, 2011 and 2013.

|  | 2011 | | 2013 | |
| --- | --- | --- | --- | --- |
| Country "A" | P(A\|A) | Rank | P(A\|A) | Rank |
| Canada | 0.97 | 1 | 0.974 | 9 |
| United States | 0.94 | 2 | 0.958 | 11 |
| Brazil | 0.93 | 3 | 0.999 | 2 |
| Malaysia | 0.92 | 4 | 0.777 | 23 |
| Poland | 0.86 | 5 | 0.936 | 14 |
| Russia | 0.86 | 6 | 0.919 | 17 |
| Great Britain | 0.86 | 7 | 0.892 | 21 |
| Norway | 0.83 | 8 | 0.917 | 18 |
| Ireland | 0.77 | 9 | 0.430 | 31 |
| Hong Kong | 0.74 | 10 | 0.626 | 28 |
| Spain | 0.67 | 11 | 0.986 | 6 |
| Jersey | 0.67 | 12 | 0.200 | 35 |
| Portugal | 0.63 | 13 | 0.985 | 7 |
| Czech Republic | 0.62 | 14 | 0.902 | 20 |
| Sweden | 0.61 | 15 | 0.867 | 22 |
| China | 0.56 | 16 | 1.000 | 1 |
| Austria | 0.51 | 17 | 0.914 | 19 |
| Germany | 0.50 | 18 | 0.688 | 26 |
| Trinidad & Tobago | 0.50 | 19 | 0.375 | 32 |
| France | 0.42 | 20 | 0.926 | 15 |
| Belgium | 0.35 | 21 | 0.968 | 10 |
| Italy | 0.32 | 22 | 0.976 | 8 |
| Singapore | 0.29 | 23 | 0.748 | 25 |
| Netherlands | 0.28 | 24 | 0.648 | 27 |
| Japan | 0.25 | 25 | 0.942 | 12 |
| Luxembourg | 0.20 | 26 | 0.590 | 29 |
| South Africa | 0.19 | 27 | 0.926 | 16 |
| Denmark | 0.13 | 28 | 0.750 | 24 |
| Bermuda | 0.11 | 29 | 0.209 | 34 |
| Switzerland | 0.11 | 30 | 0.456 | 30 |
| Mexico |  |  | 0.997 | 3 |
| India |  |  | 0.995 | 4 |
| Kenya |  |  | 0.345 | 33 |
| Argentina |  |  | 0.987 | 5 |
| Australia |  |  | 0.940 | 13 |

Notes to Table 3. This table summarizes the "in" transition statistics with respect to country labels. That is, for any country A consider all the entities incorporated in A. Here we list the fraction of the children of such an entity that are also incorporated in A (P(A|A)) for any country where in 2011 the in-country probability is neither zero or one, along with some additional countries from 2013.

Table 4. Perfect Tree Statistics (Country)

| | **May 26, 2011** | | | | **February 25, 2013** | | | |
|---|---|---|---|---|---|---|---|---|
| | Perfect | Mean | Stdev | Quantile | Perfect | Mean | Stdev | Quantile |
| **SIFIs** | | | | | | | | |
| S1 | 0.776 | 0.663 | 0.354 | 29.60% | 0.850 | 0.512 | 0.273 | 86.40% |
| S2 | 0.351 | 0.363 | 0.330 | 58.00% | 0.609 | 0.568 | 0.254 | 48.00% |
| S3 | 0.526 | 0.378 | 0.207 | 75.45% | 0.727 | 0.509 | 0.221 | 86.30% |
| S4 | 0.800 | 0.587 | 0.368 | 51.70% | 0.987 | 0.952 | 0.090 | 100% |
| S5 | 0.529 | 0.274 | 0.184 | 97.70% | 0.964 | 0.861 | 0.217 | 100% |
| S6 | 0.651 | 0.399 | 0.384 | 66.50% | 0.939 | 0.788 | 0.215 | 100% |
| S7 | 0.906 | 0.782 | 0.261 | 99.80% | 0.834 | 0.581 | 0.228 | 92.20% |
| S8 | 0.869 | 0.652 | 0.312 | 58.00% | 0.953 | 0.727 | 0.207 | 98.20% |
| S9 | 0.895 | 0.827 | 0.278 | 33.10% | 0.953 | 0.875 | 0.201 | 41.85% |
| S10 | 0.261 | 0.256 | 0.127 | 55.65% | 0.925 | 0.713 | 0.218 | 100% |
| S11 | 0.279 | 0.093 | 0.049 | 99.85% | 0.939 | 0.784 | 0.240 | 100% |
| S12 | 0.264 | 0.099 | 0.080 | 94.50% | 0.942 | 0.760 | 0.265 | 100% |
| S13 | 0.857 | 0.681 | 0.347 | 37.00% | 0.951 | 0.799 | 0.214 | 100% |
| S14 | 0.864 | 0.771 | 0.319 | 27.30% | 0.908 | 0.530 | 0.242 | 100% |
| S15 | 0.895 | 0.647 | 0.239 | 100% | 0.924 | 0.697 | 0.238 | 99.10% |
| S16 | 0.860 | 0.670 | 0.211 | 100% | 0.826 | 0.538 | 0.247 | 99.50% |
| S17 | 0.575 | 0.349 | 0.198 | 90.50% | 0.856 | 0.450 | 0.154 | 100% |
| S18 | 0.963 | 0.884 | 0.188 | 68.50% | 0.973 | 0.896 | 0.207 | 100% |
| S19 | 0.933 | 0.851 | 0.165 | 100% | 0.935 | 0.733 | 0.294 | 100% |
| *Mean* | *0.687* | *0.538* | | | *0.891* | *0.699* | | |
| **Non-SIFIs** | | | | | | | | |
| B1 | 0.962 | 0.906 | 0.120 | 99.55% | 0.959 | 0.889 | 0.150 | 100% |
| B2 | 0.941 | 0.838 | 0.248 | 36.40% | 0.991 | 0.955 | 0.095 | 100% |
| B3 | 0.520 | 0.157 | 0.136 | 100% | 0.510 | 0.152 | 0.134 | 96.60% |
| B4 | 0.669 | 0.378 | 0.238 | 100% | 0.990 | 0.960 | 0.120 | 99.90% |
| B5 | 0.429 | 0.185 | 0.086 | 100% | 0.783 | 0.422 | 0.241 | 100% |
| *Mean* | *0.704* | *0.493* | | | *0.847* | *0.676* | | |
| **Insurance Companies** | | | | | | | | |
| I1 | 0.665 | 0.548 | 0.368 | 39.30% | 0.862 | 0.581 | 0.229 | 91.55% |
| I2 | 0.297 | 0.173 | 0.169 | 65.40% | 0.752 | 0.401 | 0.190 | 99.80% |
| I3 | 0.576 | 0.308 | 0.288 | 74.80% | 0.744 | 0.379 | 0.167 | 100% |
| I4 | 0.857 | 0.735 | 0.326 | 33.20% | 0.884 | 0.603 | 0.232 | 90.90% |
| I5 | 0.781 | 0.615 | 0.253 | 63.15% | 0.909 | 0.550 | 0.201 | 99.90% |
| *Mean* | *0.635* | *0.476* | | | *0.830* | *0.503* | | |

Notes to Table 4: Summary statistics describing the regularity of a firm's country control hierarchy for two periods in time, May 26, 2011 (left panel) and February 25, 2013 (right panel). The columns labeled "Perfect" refer to the tree-similarity statistic equal to the number of nodes that have the same country designation as their immediate parent divided by the total number of nodes in the firm's tree. In a perfect tree, deletion of a node and its children will not change the value of the statistic. From a supervisory perspective, this means that supervisory oversight follows firm organizational lines. A number closer to one signifies greater proximity to perfect. The columns "Mean", "Stdev" and "Quantile" are summary statistics describing simulation with 1000 replications where each firm's tree structure is taken as given and the node identifiers are randomly drawn according tothe actual probability distribution of the firm.

Table 5.  Perfect Tree Statistics (SIC)

| | **May 26, 2011** | | | | **February 25, 2013** | | | |
|---|---|---|---|---|---|---|---|---|
| | Perfect | Mean | Stdev | Quantile | Perfect | Mean | Stdev | Quantile |
| **SIFIs** | | | | | | | | |
| S1 | 0.305 | 0.065 | 0.067 | 100% | 0.631 | 0.377 | 0.157 | 99.30% |
| S2 | 0.103 | 0.061 | 0.035 | 83.40% | 0.466 | 0.647 | 0.202 | 20.80% |
| S3 | 0.092 | 0.051 | 0.039 | 90.80% | 0.575 | 0.358 | 0.164 | 100% |
| S4 | 0.561 | 0.690 | 0.310 | 30.90% | 0.955 | 0.925 | 0.113 | 32.55% |
| S5 | 0.089 | 0.061 | 0.064 | 81.70% | 0.940 | 0.885 | 0.172 | 53.65% |
| S6 | 0.435 | 0.307 | 0.316 | 62.40% | 0.830 | 0.731 | 0.194 | 50.35% |
| S7 | 0.022 | 0.090 | 0.027 | 5.80% | 0.601 | 0.389 | 0.194 | 77.85% |
| S8 | 0.473 | 0.460 | 0.273 | 47.80% | 0.754 | 0.655 | 0.183 | 63.60% |
| S9 | 0.359 | 0.358 | 0.193 | 43.00% | 0.841 | 0.670 | 0.261 | 55.15% |
| S10 | 0.032 | 0.044 | 0.025 | 40.65% | 0.767 | 0.594 | 0.251 | 60.20% |
| S11 | 0.465 | 0.268 | 0.107 | 95.20% | 0.938 | 0.873 | 0.126 | 90.80% |
| S12 | 0.170 | 0.271 | 0.162 | 34.40% | 0.820 | 0.741 | 0.240 | 43.40% |
| S13 | 0.856 | 0.558 | 0.314 | 95.25% | 0.879 | 0.790 | 0.188 | 50.90% |
| S14 | 0.078 | 0.114 | 0.100 | 51.40% | 0.706 | 0.412 | 0.266 | 75.30% |
| S15 | 0.683 | 0.489 | 0.213 | 81.90% | 0.784 | 0.586 | 0.175 | 92.80% |
| S16 | 0.417 | 0.260 | 0.168 | 70.20% | 0.491 | 0.530 | 0.149 | 35.40% |
| S17 | 0.291 | 0.151 | 0.117 | 86.85% | 0.471 | 0.213 | 0.057 | 100% |
| S18 | 0.666 | 0.495 | 0.204 | 75.10% | 0.844 | 0.736 | 0.235 | 37.30% |
| S19 | 0.836 | 0.664 | 0.237 | 74.90% | 0.755 | 0.427 | 0.266 | 92.20% |
| *Mean* | *0.365* | *0.287* | | | *0.739* | *0.607* | | |
| **Non-SIFIs** | | | | | | | | |
| B1 | 0.754 | 0.646 | 0.260 | 54.20% | 0.871 | 0.755 | 0.243 | 50.95% |
| B2 | 0.621 | 0.349 | 0.303 | 68.80% | 0.968 | 0.930 | 0.113 | 66.55% |
| B3 | 0.276 | 0.116 | 0.078 | 97.05% | 0.671 | 0.447 | 0.258 | 67.35% |
| B4 | 0.103 | 0.038 | 0.034 | 89.75% | 0.957 | 0.935 | 0.158 | 14.40% |
| B5 | 0.302 | 0.214 | 0.111 | 71.00% | 0.659 | 0.529 | 0.288 | 48.70% |
| *Mean* | *0.411* | *0.273* | | | *0.825* | *0.719* | | |
| **Insurance Companies** | | | | | | | | |
| I1 | 0.375 | 0.442 | 0.295 | 42.90% | 0.546 | 0.366 | 0.167 | 86.60% |
| I2 | 0.212 | 0.115 | 0.044 | 97.35% | 0.488 | 0.274 | 0.084 | 99.50% |
| I3 | 0.125 | 0.067 | 0.046 | 85.95% | 0.596 | 0.440 | 0.206 | 62.80% |
| I4 | 0.478 | 0.420 | 0.248 | 54.00% | 0.497 | 0.403 | 0.125 | 72.10% |
| I5 | 0.119 | 0.055 | 0.017 | 99.70% | 0.564 | 0.369 | 0.110 | 97.50% |
| *Mean* | *0.262* | *0.220* | | | *0.538* | *0.370* | | |

Notes to Table 5: Summary statistics describing the regularity of a firm's SIC control hierarchy for two periods in time, May 26, 2011 (left panel) and February 25, 2013 (right panel).  The columns labeled "Perfect" refer to the tree-similarity statistic equal to the number of nodes that have the same SIC designation as their immediate parent divided by the total number of nodes in the firm's tree.   In a perfect tree, deletion of a node and its children will not change the value of the statistic.  From a supervisory perspective, this means that supervisory oversight follows firm organizational lines.  A number closer to one signifies greater proximity to perfect.   The columns "Mean", "Stdev" and "Quantile" are summary statistics describing simulation with 1000 replications where each firm's tree structure is taken as given and the node identifiers are randomly drawn according tothe actual probability distribution of the firm.

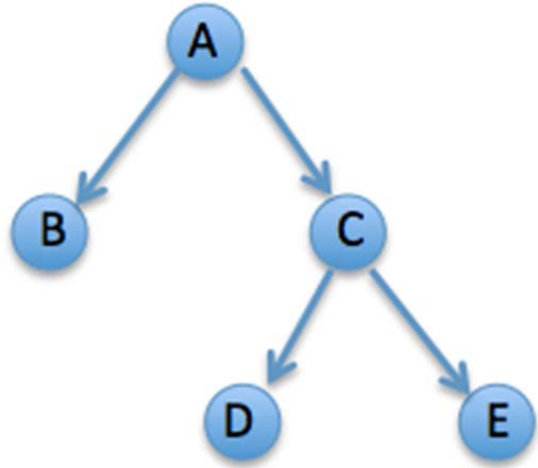

Figure 1. Basic network terminology and structures. This figure shows a particular example of rooted directed tree. The root is node A, while B, D, and E are leaves or leaf nodes. Node C is neither the root nor a leaf and is sometimes called an internal node. Nodes B and C are children of the node A, which is the parent of these nodes. In addition, this is a regularly branching tree in which each node that has children has exactly two children. This tree has depth 2 (the distance of from node A to either node D or node E) and a total of five nodes. If this were the tree corresponding to a control structure of a financial institution, then the "ultimate parent" would be node A and nodes B and C would be direct subsidiaries of A in which A still held a controlling interest, while D and E would denote subsidiaries of C in which C held a controlling interest.

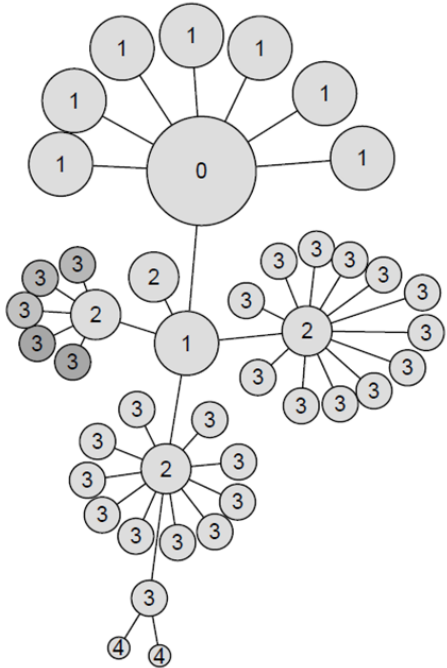
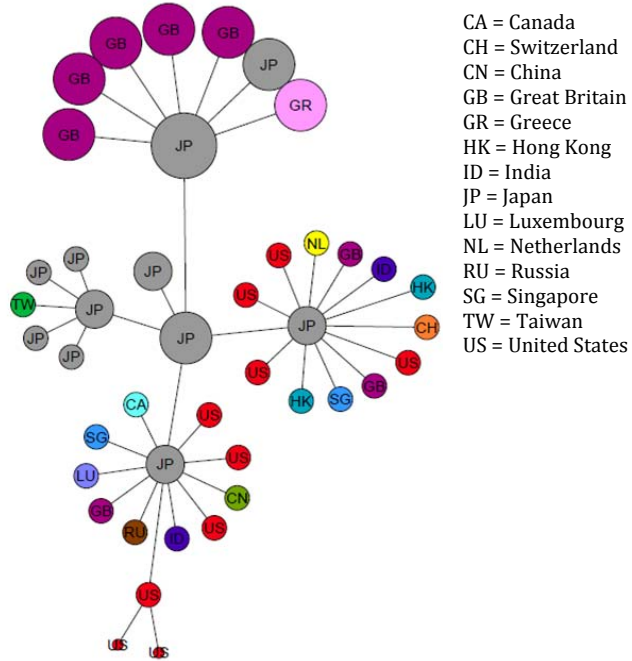
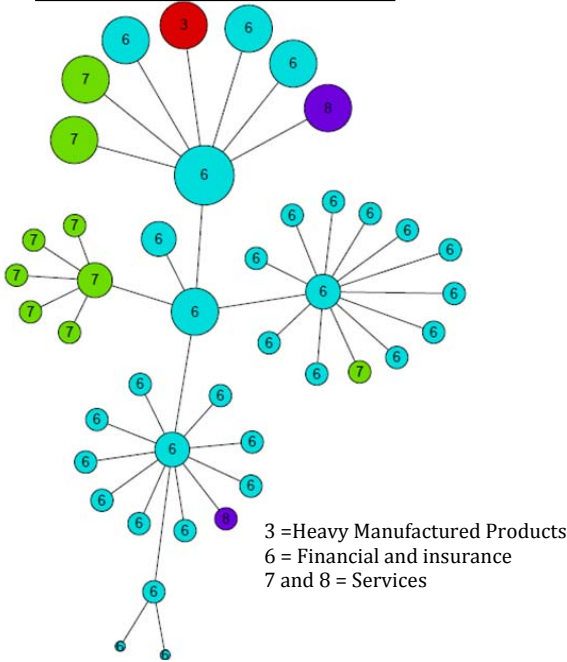
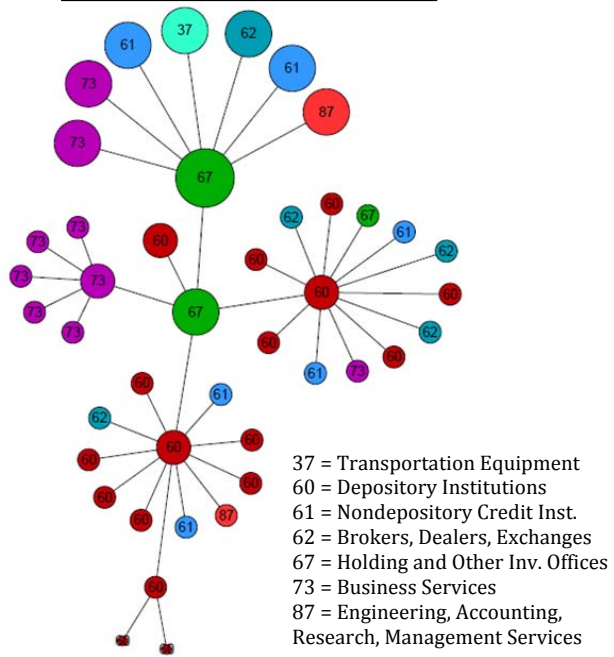

Figures 2a-d. The control hierarchy of SIFI S11, color-coded by distance from the ultimate parent, by depth, country, and both 1-digit and 2-digit SIC classifications (classifications are available at www.secinfo.com/$/SEC/SIC.asp?Division=I). A consistent layout is used for all three representations, for comparability. Node size is proportional to distance to the ultimate parent, with larger nodes corresponding to closer distance.

Figure 3a: by depth (n=5)

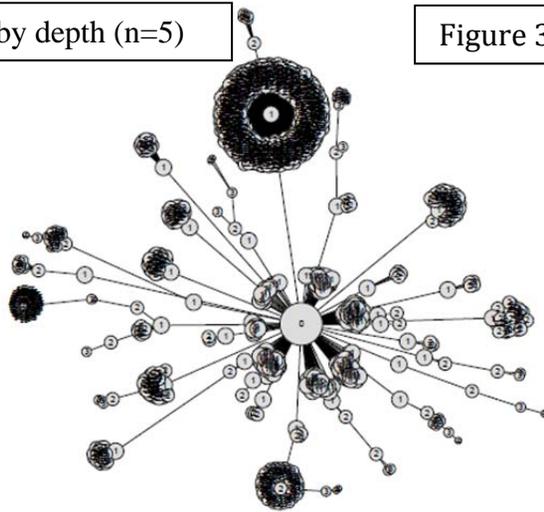

Figure 3b: by country (n=32)

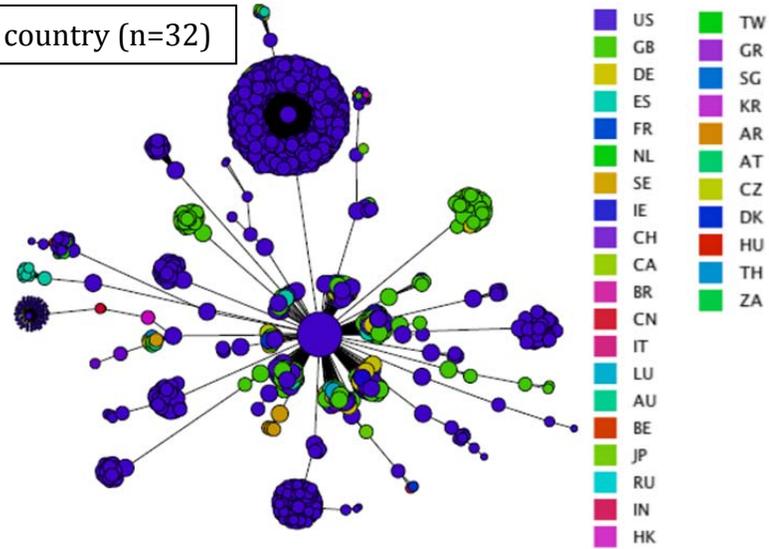

Figure 3c: by 1-digit SIC classification (n=100)

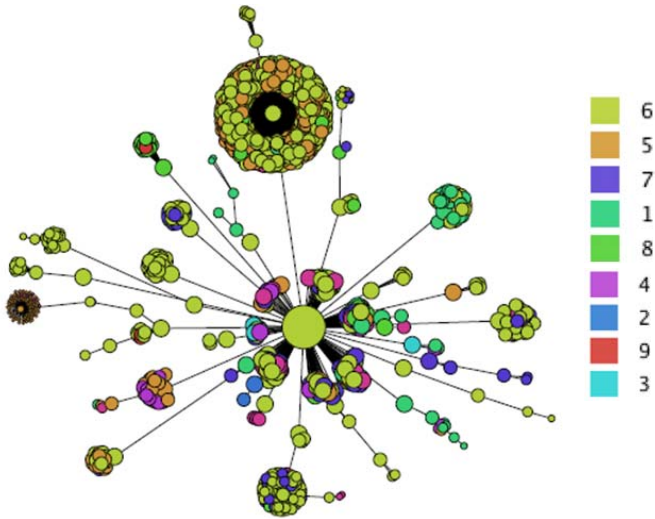

Figures 3a-c. The control hierarchy of SIFI S16, color-coded by distance from the ultimate parent, by depth, country, and 1-digit SIC classification. A consistent layout is used for all three representations, for comparability. Node size is proportional to distance to the ultimate parent, with larger nodes corresponding to closer distance.

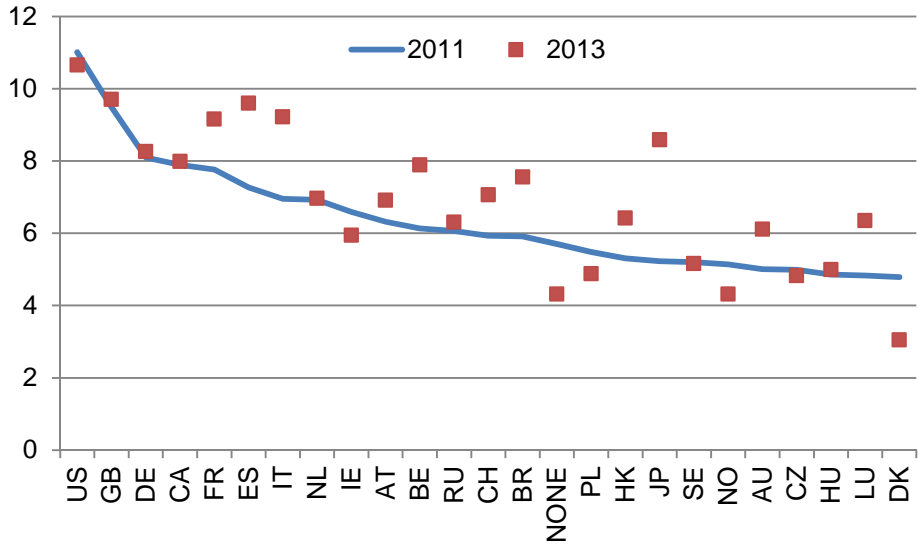

Figure 4. Country distribution of entities (log-scale), top 25 countries, 2011 and 2013. The countries are listed in declining order based on their 2011 distribution. Because the order of the countries changes between years, the 2013 distribution is represented by the markers. This presentation therefore also shows how each country's ranking changed during the two years. "NONE" denotes no classification was available.

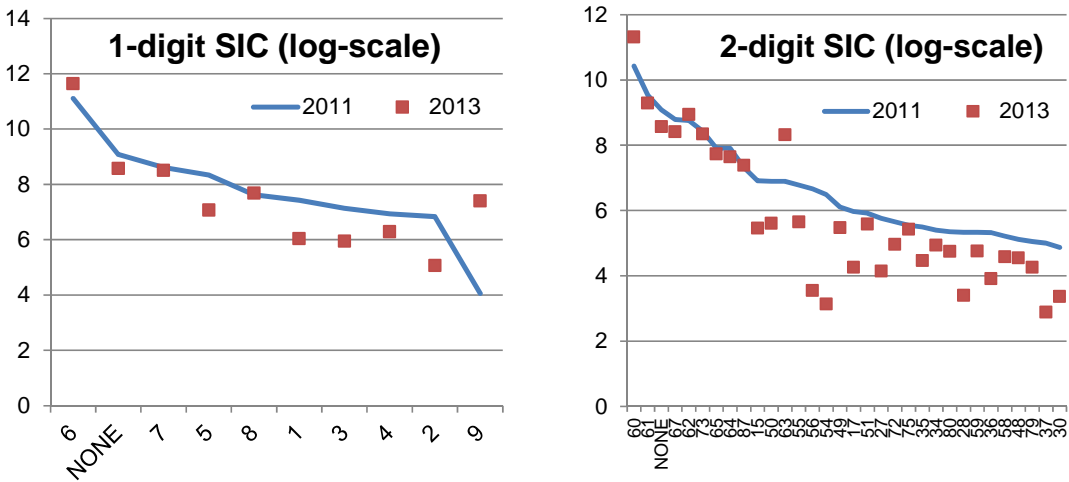

Figure 5. SIC distribution of entities (log scale), 2011 and 2013. On the left we see the SIC distribution at a resolution of the first digit and on the right we see it to the second digit. "NONE" denotes no classification was available. The codes are listed in declining order of representation based on their 2011 distribution. Because the ranking of SIC codes (in terms of number of nodes) changes between years, the 2013 distribution is represented by the markers. This presentation therefore also shows how each SIC code's ranking changed during the two years.

**2011**

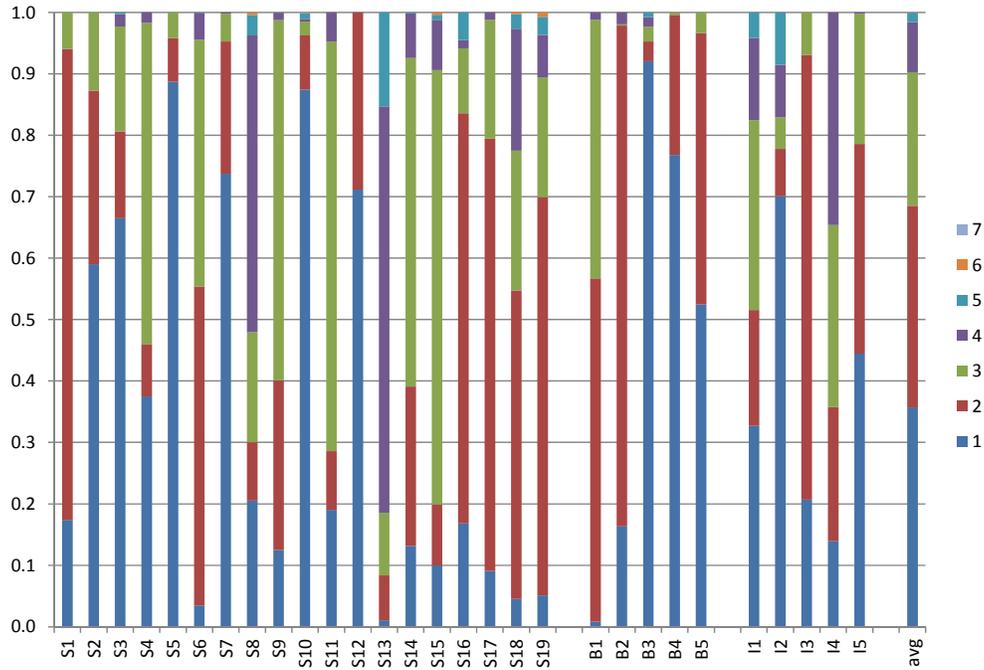

**2013**

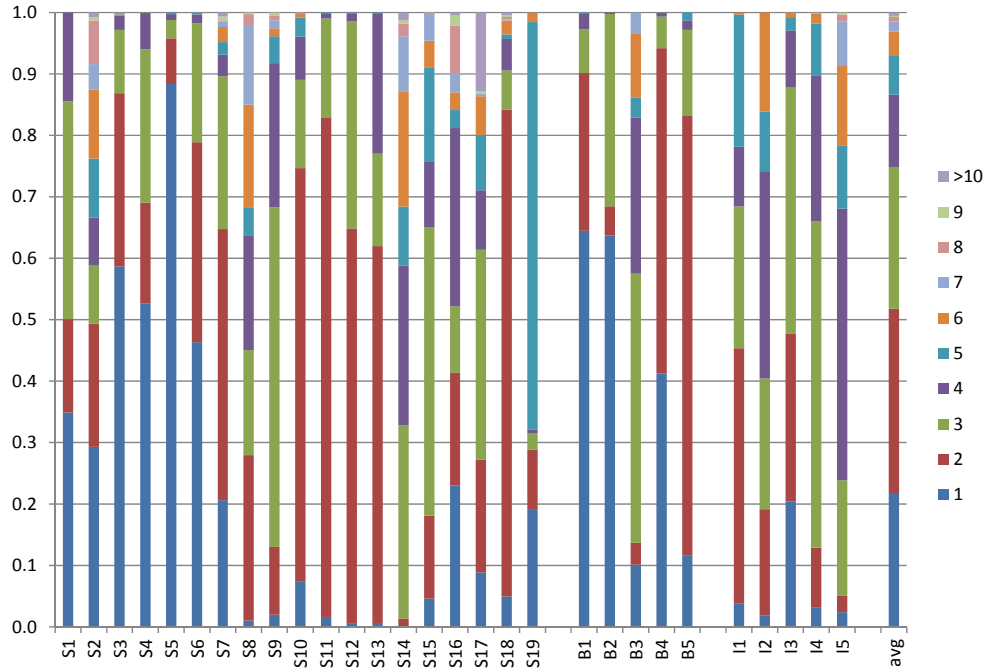

Figure 6. For each firm, we plot the fraction of nodes at each level in its control hierarchy (i.e., the distance from the root), for 2011 and 2013. For ease of comparability between 2011 and 2013, for firms with more than nine levels in 2013, the fraction of nodes beyond level 9 are aggregated into the ">10" distance. The fraction of nodes at each level for the entire sample of firms we consider is shown in the right-most column ("avg").

Figure 7. Scatter plot of firms size rank versus country tree similarity rank

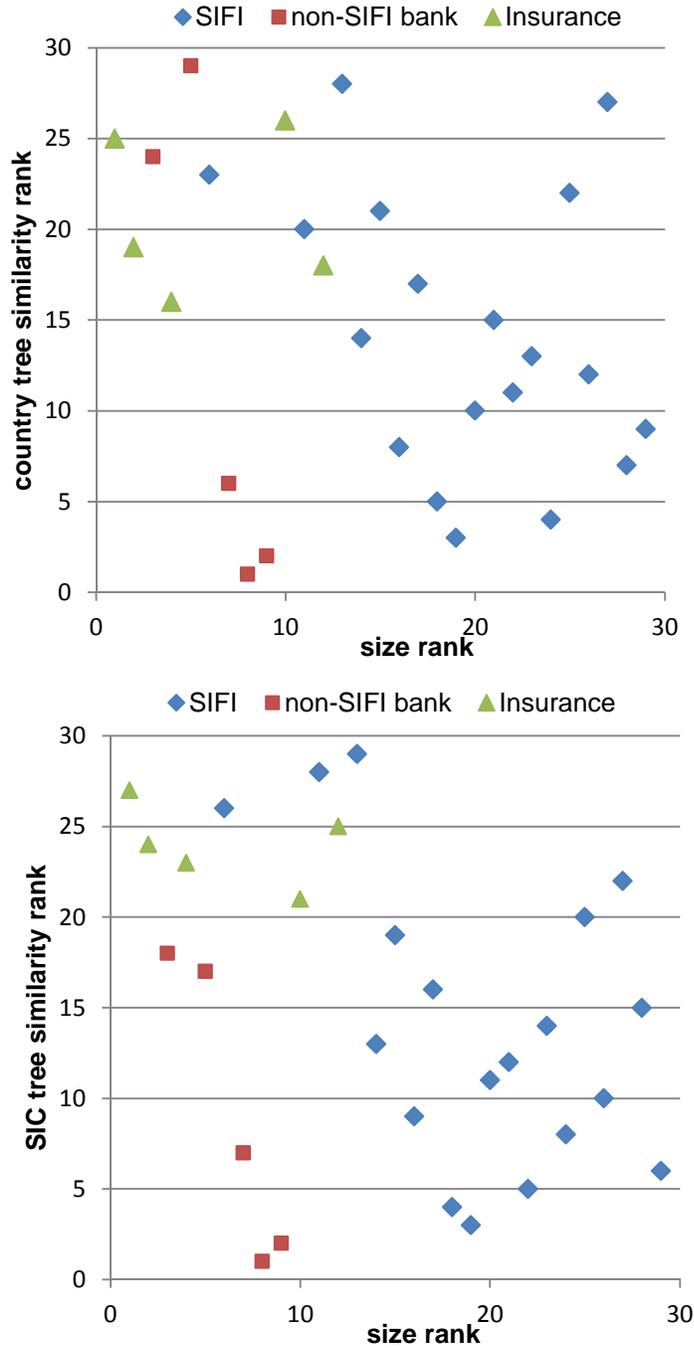

Notes: These charts show the distribution of size rank versus 2013 country (top chart) and SIC (bottom chart) tree similarity ranks, for all 29 firms in our sample, shown separately for SIFIs, non-SIFI banks, and insurance companies. A higher rank indicates greater complexity and/or larger size. While it is evident that the non-SIFI banks and insurance companies are smaller in size than the SIFIs, there is also evidence that the insurance companies have country tree similarity statistics that are farther from a perfect tree (i.e., they are more intermingled and hence potentially more difficult to supervise/regulate). In contrast, the non-SIFI banks are similar to their SIFI counterparts in that some have tree structures that are close to perfect and others far away. Data on size as of December 31, 2012 are obtained from Bloomberg®.

Appendix A: List of Financial Institutions

This is the list of financial institutions analyzed in this paper, broken out into banks and insurance companies and grouped by country of incorporation. Those that are among the twenty-nine systemically important financial institutions (SIFIs), as determined by the FSB and IMF, are preceded by an asterisk. The country of incorporation is included in parentheses. For completeness, the SIFIs that are not included in our dataset are also listed.

| **Banks** | **Insurance Companies** |
|---|---|
| *Bank of America (US) | Allianz (DE) |
| *Citigroup (US) | Aviva (GB) |
| *Goldman Sachs (US) | Axa (FR) |
| *JP Morgan Chase (US) | Swiss Re (CH) |
| *Morgan Stanley (US) | Zurich (CH) |
| Royal Bank of Canada (CA) | |
| *Barclays PLC (GB) | |
| *HSBC Holdings PLC (GB) | |
| *Royal Bank of Scotland PLC (GB) | |
| Standard Chartered (GB) | |
| *Credit Suisse AG (CH) | |
| *UBS AG (CH) | |
| *BNP Paribas SA (FR) | |
| *Société Générale SA (FR) | |
| BBVA (ES) | |
| *Banco Santander SA (ES) | |
| *Mitsubishi UFJ FG (JP) | |
| *Mizuho FG (JP) | |
| Nomura (JP) | |
| *Sumitomo Mitsui FG (JP) | |
| Banca Intesa (IT) | |
| *UniCredit (IT)* | |
| *Deutsche Bank AG (DE) | |
| *ING Groep NV (NL) | |

SIFIs not included in the dataset:
[Wells Fargo (US) *]
[Lloyds (GB)*]
[Banque Populaire (FR)*]
[Crédit Agricole (FR)*]
[Commerzbank (DE)*]
[Dexia (BE) *]
[Bank of China (CN)*]
[Nordea (SW) *]